\documentclass[floats,floatfix,showpacs,amssymb,prd,twocolumn,superscriptaddress,nofootinbib]{revtex4-1}

\usepackage{amssymb,amsmath,verbatim,mathtools,needspace,enumitem,etoolbox,graphicx,physics,microtype}

\usepackage[dvipsnames, usenames]{xcolor}

\definecolor{linkcolor}{rgb}{0.0,0.3,0.5}
\usepackage[unicode, colorlinks=true, linkcolor=linkcolor, citecolor=linkcolor, filecolor=linkcolor,urlcolor=linkcolor, pdfusetitle]{hyperref}
\usepackage[all]{hypcap}
\usepackage[T1]{fontenc}
\usepackage[utf8]{inputenc}

\allowdisplaybreaks
\graphicspath{{./figure/}}

\begin{document}

\title{Expanding the LISA Horizon from the Ground}

\author{Kaze W.~K. Wong}
\affiliation{Department of Physics and Astronomy, Johns Hopkins University, Baltimore, MD 21218 USA}
\author{Ely D. Kovetz}
\affiliation{Department of Physics and Astronomy, Johns Hopkins University, Baltimore, MD 21218 USA}
\author{Curt Cutler}
\affiliation{Theoretical Astrophysics, California Institute of Technology, Pasadena, California 91125, USA}
\affiliation{Jet Propulsion Laboratory,  4800 Oak Grove Drive, Pasadena, CA 91109, USA}
\author{Emanuele Berti}
\affiliation{Department of Physics and Astronomy, Johns Hopkins University, Baltimore, MD 21218 USA}
\affiliation{Department of Physics and Astronomy, The University of Mississippi, University, MS 38677, USA}

\begin{abstract}

The Laser Interferometer Space Antenna (LISA) gravitational-wave (GW) observatory will be  limited in its ability to detect mergers of binary black holes (BBHs) in the stellar-mass range. A future ground-based detector network, meanwhile, will achieve by the LISA launch date a sensitivity that ensures complete detection of all mergers within a volume $>\!\mathcal{O}(10)\,{\rm Gpc}^{3}$.
We propose a method to use the information from the ground to revisit the LISA data in search for sub-threshold events. By discarding spurious triggers that do not overlap with the ground-based catalogue, we show that the signal-to-noise threshold $\rho_{\rm LISA}$ employed in LISA can be significantly lowered, greatly boosting the detection rate. The efficiency of this method depends predominantly on the rate of false-alarm increase when the threshold is lowered and on the uncertainty in the parameter estimation for the LISA events. As an example, we demonstrate that while all current LIGO BBH-merger detections would have evaded detection by LISA when employing a standard $\rho_{\rm LISA}=8$ threshold, this method will allow us to easily (possibly) detect an event similar to GW150914 (GW170814) in LISA. Overall, we estimate that the total rate of stellar-mass BBH mergers detected by LISA can be boosted by a factor $\sim\!4$ ($\gtrsim\!8$) under conservative (optimistic) assumptions. 
This will enable new tests using multi-band GW observations, significantly aided by the greatly increased lever arm in frequency.
\end{abstract}

\maketitle

Multi-band measurements of GWs~\cite{Sesana:2016ljz} from coalescing binary black holes (BBHs) can open the door to a wide array of invaluable studies. Spanning a wider range of frequencies will increase sensitivity to eccentric orbits, which can be used to distinguish between different binary formation channels, improve merger-rate estimation, allow for more precise tests of gravity and assist in instrument calibration. 
Better science will be enabled if many events are detected in both a ground-based network (Ground) and a space observatory such as LISA.

Unfortunately, LISA will not be nearly as sensitive as the Ground detectors to stellar-mass BBH mergers. This issue affects in particular ``multiband'' inspiral events, for which the GW frequency drifts from the LISA to the Ground band during the LISA observation window. 
This condition determines a minimum frequency at which the event can appear in LISA (typically $\gtrsim{10}^{-2}\, \rm{Hz}$ for stellar-mass BBHs). 
Taking advanced LIGO (aLIGO) at design sensitivity as an example and adopting a similar signal-to-noise threshold of $\rho=8$ in both experiments, the fraction of aLIGO events that will be detectable in LISA is less than $1\%$.

If we can manage to lower the LISA signal-to-noise threshold, the horizon distance (which is the maximum distance at which a source is detectable) will grow, and the increase in accessible volume will result in a rapid rise in the multi-band detection rate.
Setting a lower threshold, however, means that we increase the risk of classifying noise triggers as real events (false alarms). The false-alarm rate (FAR) is a steep function of $\rho$~\cite{Capano:2016uif}.

In this {\it Letter} we propose a method to discard spurious LISA triggers that show up as the signal-to-noise threshold is lowered, using information from the Ground. We show that a large number of random noise triggers can be filtered out by imposing consistency with Ground measurements for multiple parameters in tandem. 

The procedure is as follows: we first set an initial threshold, e.g.~$\rho_{\rm LISA}\!=\!8$, and determine which (real) events in the Ground catalogue are detectable in LISA with this threshold. The parameters of all LISA candidate events identified with this threshold are then compared with those in the Ground list (taking into account the LISA parameter-estimation uncertainty), and those that do not overlap with any real event are discarded. We lower the threshold and iterate this procedure until the probability that a random  trigger is consistent with some Ground event becomes significant.

\begin{figure}
\includegraphics[width=\columnwidth]{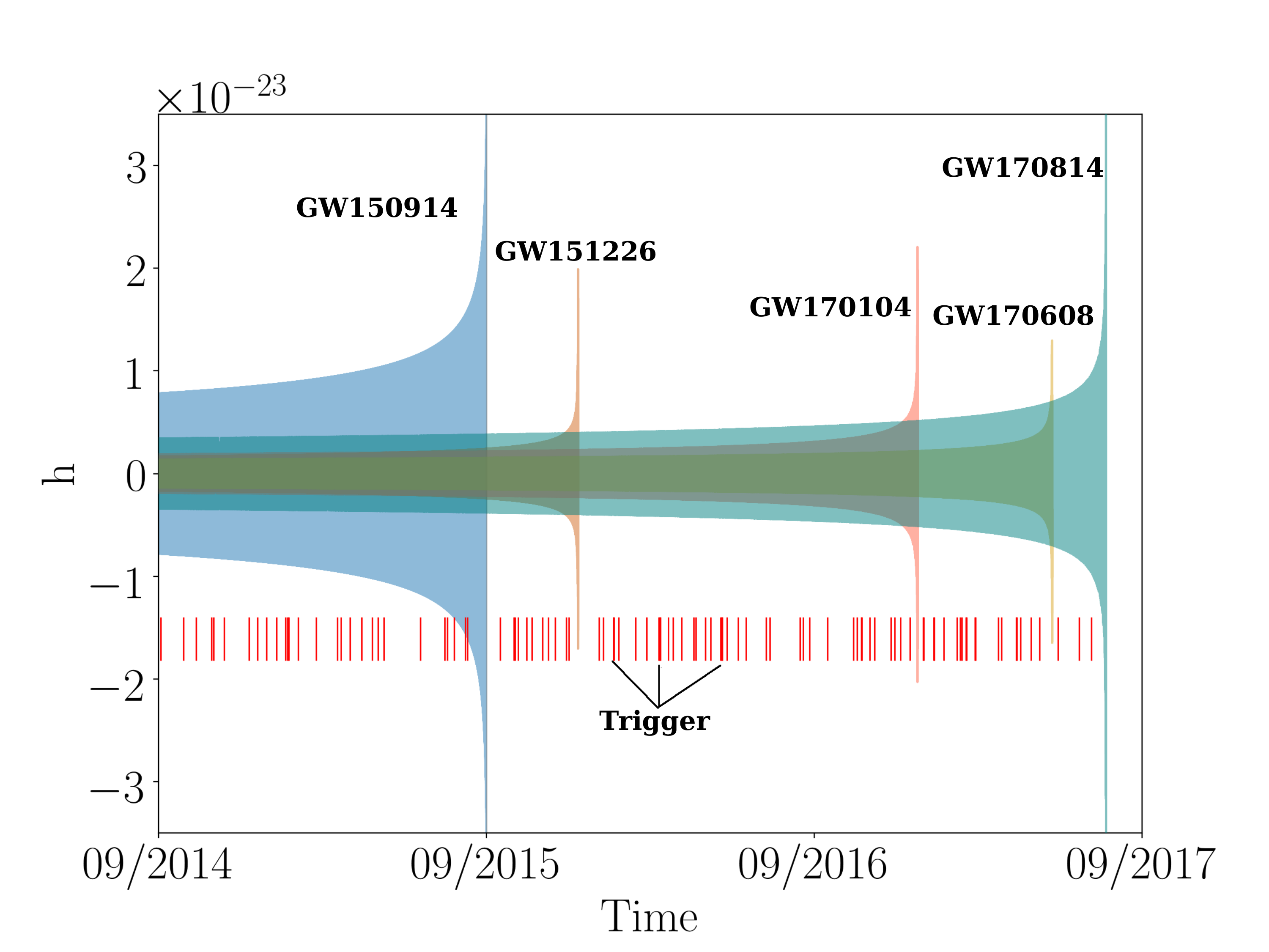}
\caption{
Illustration of our method to discard LISA triggers.
The waveforms are those of the gravitational events which were observed by aLIGO in its O1 and O2 runs (2015-2017).  GW150914 would have had the highest signal-to-noise in LISA, $\rho_{\rm LISA}\!=\!7$, while GW170814 would have had $\rho_{\rm LISA}\!=\!4.5$ (assuming 4 years of integration time), both of which are below the conventional $\rho\!=\!8$ threshold. The red stripes indicate the merger time of LISA triggers (their width set by the uncertainty). If a trigger does not agree with any of the events detected from the Ground, it can be discarded as random noise (or as an astrophysical event whose merger will appear in LIGO in the future and is thus irrelevant for our purposes).
We show that if LISA had started observing in 2011, it would have been possible to lower its signal-to-noise threshold and recover GW150914, and potentially also GW170814. The other events would have been out of reach.
}
\label{fig:waveforms}
\end{figure}
Figure \ref{fig:waveforms} illustrates the concept of filtering spurious triggers using only $t_c$, the time of coalescence, as the discarding parameter. Compared with the entire LISA observation time, $\mathcal{O}(1)\,{\rm years}$, the typical uncertainty on $t_c$ as determined by LISA is $\sim 7$ orders of magnitude smaller, $\mathcal{O}(10)\,{\rm seconds}$. With $\mathcal{O}(1000)$ events expected to be detected from the Ground within the volume accessible by LISA with $\rho_{\rm LISA}\!\gtrsim\!5$, we should therefore be able to filter out roughly $\gtrsim\!10^4$ random triggers based on $t_c$ alone. This will allow a detection of events with $\rho_{\rm LISA}\!\sim\!7$, such as GW150914~\cite{Abbott:2016blz}, over the LISA mission lifetime. We will see that incorporating additional parameters may enable a multi-band detection of events with $\rho_{\rm LISA}\sim4$, such as GW170814~\cite{Abbott:2017gyy}.

In what follows we choose to focus on three  waveform ingredients: the source masses, sky location and merger time.
We will test the efficiency of our proposed method based on a Fisher matrix analysis to estimate the parameter estimation uncertainty in the LISA band~\cite{Berti:2004bd}, and report the potential improvement in the LISA event rate given different assumptions about the FAR and the BBH mass function. 

We assume the posteriors to be Gaussian, so a trigger is characterized by its $k$-dimensional vector of best-fit parameter values $\vec{\mu}$ and covariance matrix $\Sigma$. The problem of consistency checking between the LISA and Ground measurements corresponds to finding the overlap between two volumes in a multi-dimensional space given some metric. We  claim that two measurements taken by LISA and the Ground agree with each other if they meet the following criterion:
\begin{align}
D({\vec{\mu}}_{\textrm{LISA}},{\vec{\mu}}_{\textrm{Ground}},{\Sigma}_{\textrm{LISA}},{\Sigma}_{\textrm{Ground}}) \leq {\chi}^{2}_{k}(p),
\label{criteria}
\end{align}
where $D$ is a function that gives the distance between two points in the high-dimensional space under some metric, and
${\chi}^{2}_{k}(p)$ is the quantile function for probability $p$ of the Chi-Squared distribution with $k$ degrees of freedom.

A typical source in LISA will be characterized by $k=9$ parameters (when taking into account the antenna pattern),
so the exact two-point distance problem would be solved in an $18$-dimensional space, and hence it can be computationally intensive.
Instead of solving the problem exactly,
we calculate the volume bounded by ${\chi}^{2}_{k}(p)$ in the parameter space centered at the best-fit value for each parameter that is given by the more precise measurement between the Ground and LISA. Since
most of the sources will be detected from the Ground with signal-to-noise well above threshold, the Ground measurements can be treated as the ``true" values (neglecting any systematic bias).
The consistent volume in parameter space of a particular source with parameters $\vec{\theta}$ will be well-approximated by the ellipsoid 
\begin{align}
V(\vec{\theta},p) = {\left(\frac{{\chi}^{2}_{k}(p)}{{\chi}^{2}_{k}(0.67)}\right)}^{k}\sqrt{{(2\pi)}^{k}|\Sigma_{\rm LISA}(\vec{\theta})|},
\label{eq:Volume}
\end{align}
where $|\Sigma_{\rm LISA}(\vec{\theta})|$ denotes the determinant of the covariance matrix given by LISA
using the most recent noise power spectral density ${S}_{n}(f)$~\cite{Cornish:2018dyw},
and ${\chi}^{2}_{k}(0.67)$ corresponds to a bound at "$1\sigma$" level.
The fraction of triggers which are consistent between the two detectors is then given by
\begin{align}
  {f}_{c}(\rho,T) =
  \frac{\int {d{\vec{\theta}}\ {n}_{s}(\vec{\theta},\rho},T)\int_{V(\vec{\theta},p)}d\vec{{\theta}^{'}}{{n}_{b}(\vec{{\theta}^{'}})}}{\int{{n}_{b}(\vec{{\theta}^{'}})d\vec{{\theta}^{'}}}},
\label{eq:fc}
\end{align}
where ${n}_{s}(\vec{\theta},\rho,T)$ is the number of astrophysical (real) events which LISA is sensitive to (all of which are detectable from the Ground) for a given vector $\vec{\theta}$ of source parameters, a signal-to-noise threshold $\rho_{\rm LISA}$, and integration time $T$; ${n}_{b}(\vec{\theta'})$ is the number density of LISA triggers as a function of $\vec{\theta'}$ in the search parameter space.

The most important ingredient in our analysis is the 
relationship between the threshold $\rho_{\rm LISA}$ and the number of expected background triggers, which we call the ``FAR curve.'' 
At this time, there is no reliable estimate for the LISA FAR curve.
We therefore use as a proxy the results of the LIGO Mock Data Challenge~\cite{Capano:2016uif},
which suggest that the number of background triggers increases by about two orders of magnitude when the signal-to-noise threshold is decreased by one (we use their Experiment 3, which is the most relevant for our study).
This agrees with the recent findings of Ref.~\cite{Lynch:2018yom}.

We can then define the {\it effective} LISA threshold as 
\begin{align}
{\rho}^{\textrm{eff}}_{\rm LISA}(T) = {\rho}^{0}_{\rm LISA}+{\rm{log}}_{\Gamma}({f}_{c}({\rho}^{\textrm{eff}}_{\rm LISA},T)),
\label{eq:SNReff}
\end{align}
where ${\rho}^{0}_{\rm LISA}$ is the conventional signal-to-noise threshold, $\Gamma$ is the FAR and $T$ is the integration time in LISA.
Eq.~\eqref{eq:SNReff} is the crux of the method proposed in this work.

The FAR curve given in Ref.~\cite{Capano:2016uif} has a slope $\Gamma\sim100$ and is not shown below ${\rho}\! =\! 5.5$. 
As a conservative estimate, we impose an exponential cutoff ${e}^{-3(\rho-5.5)}$ starting at $\rho\!=\!5.5$, essentially preventing any improvement beyond $\rho=5$.
We also consider a more optimistic case in which we extrapolate the FAR curve with a similar cutoff at $\rho\!=\! 4$.
Given the volume permitted by a single source, Eq.~\eqref{eq:Volume}, the number density $n_s$ of real sources in the parameter space and the FAR function, we are now ready to obtain ${\rho}_{\textrm{eff}}$ by solving Eq.~\eqref{eq:SNReff} self-consistently.

In order to compute the second integral in Eq.~\eqref{eq:fc}, we need to estimate ${\Sigma}_{\rm LISA}$.
We adopt a modification of the Fisher matrix code from Ref.~\cite{Berti:2004bd} to calculate the uncertainties on source parameters. As explained above, we calculate ${\rho}^{\textrm{eff}}_{\rm LISA}(T)$ using the three groups of parameters which contribute the most to the fraction of discarded events $f_c$: \\ 
(i) Time of coalescence ${t}_{c}$: we care only for events that will merge in the Ground frequency band and assume that noise triggers will be distributed uniformly in the LISA observation window, which is determined by $T$.\\ 
(ii) Component masses $({M}_{1}$, ${M}_{2})$: we assume that noise triggers will pick up a random template in the template bank, and calculate the fraction $f_c$ assuming noise triggers are distributed uniformly in the $({M}_{1},\,{M}_{2})$ plane.
The uncertainty on either component mass is normally $\sim 10\%$ of the measured value, but due to the strong correlation between the two component masses~\cite{TheLIGOScientific:2016pea}, the allowed volume in the parameter space is typically much smaller than $10\%$. This volume is related to the uncertainty in chirp mass measurement, which is expected to be quite small in LISA (as BBHs spend many cycles in its frequency band).
Typically the probability of a noise trigger being consistent with one real event is $\sim\!10^{-6}$.\\
(iii)  Sky location $({\theta}_{S}$, ${\phi}_{S})$:
We assume that noise triggers will be uniformly distributed across the sky. LISA will be able to localize sources to within $\mathcal{O}(10)\,{\rm deg}^2$~\cite{DelPozzo:2017kme}. Comparing to the whole sky, the probability of a noise trigger being consistent with one event is $\lesssim{10}^{-3}$.

Our figure-of-merit is the number of additional sources we can recover in LISA by replacing the conventional  threshold $\rho^0_{\rm LISA}$ with $\rho_{\rm LISA}^{\rm eff}$. This of course depends on the astrophysical BBH merger rate.
Multiband events probed by LISA are in the local Universe, so we can assume the merger rate $R$ to be constant in redshift. We denote by $\Lambda$ the mean rate of events of 
astrophysical origin above a certain signal-to-noise threshold, given by
$\Lambda = R \left<VT\right>$,
where $\left<VT\right>$ is the time and population-averaged space-time volume accessible to the detector at the chosen threshold $\rho^{\rm th}$, defined as~\cite{Abbott:2016nhf}
\begin{align}
\left<VT\right> = {T}\int_{}^{}{dzd\vec{\theta}\,\frac{d{V}_{c}}{dz}\frac{1}{1+z}s(\vec{\theta})f(z,\vec{\theta},\rho^{\rm th})},
\label{eq:VT}
\end{align}
where ${V}_{c}$ is the comoving volume, $s(\vec{\theta})$ is the injected distribution of source parameters, and $0 \leq f(z,\vec{\theta},\rho^{\rm th})\leq1$ is the fraction of injections detectable by the experiment.
  
In order to calculate $\left<VT\right>$, we need to solve for the horizon distance and redshifted volume as a function of source parameters \cite{Chen:2017wpg}, and then marginalize over an input population $s(\vec{\theta})$.
We consider sources characterized by 9 parameters: the two component masses $({M}_{1},{M}_{2})$,
time of coalescence ${t}_{c}$,
phase of coalescence ${\phi}_{c}$,
luminosity distance ${D}_{L}$,
sky locations of the source $({\bar{\theta}}_{S},{\bar{\phi}}_{S})$, and the orbital angular momentum direction  $({\bar{\theta}}_{L},{\bar{\phi}}_{L})$.
In practice, we sample over the two component masses and four sky locations, with ${t}_{c}$ and ${\phi}_{c}$ arbitrarily set to zero.

For the injected mass distribution, we follow Ref.~\cite{Kovetz:2016kpi} and define the probability density function (PDF) of ${M}_{1}$ 
\begin{align}
\mathit{P}({M}_{1}) \equiv {A}_{{M}_{1}}{{M}_{1}}^{-\alpha}\mathcal{H}({M}_{1}-{M}_{\rm gap}){e}^{-{({M}_{1}/{M}_{\rm cut})}^{2}},
\end{align}
where ${A}_{{M}_{1}}$ is a normalization constant, $\mathcal{H}$ is the Heaviside function,
${M}_{\rm gap}$ is the minimum mass of a stellar black hole (assumed to be $5 {M}_{\odot}$), and
by default we set the upper cutoff ${M}_{\rm cut}=40{M}_{\odot}$~\cite{Belczynski:2016jno,Fishbach:2017zga,Kovetz:2018vly}.
To account for uncertainty regarding these choices, we also calculate our results using two other mass functions:
in one we replace the Gaussian cutoff with a sharp step function $P(M)\propto\mathcal{H}({M}_{\odot}-M_{\rm cut})$,
and in another with an exponential cutoff $P(M)\propto e^{-M_1/M_{\rm cut}}$. 
For all cases we limit the maximum component mass to $100{M}_{\odot}$.
Finally, given a value for ${M}_{1}$, we define the PDF of ${M}_{2}$ as a uniform distribution ranging from ${M}_{\textrm{gap}}$ to ${M}_{1}$~\cite{TheLIGOScientific:2016pea,Kovetz:2016kpi}:
\begin{align}
\mathit{P}({M}_{2}\mid{M}_{1}) \equiv {A}_{{M}_{2}}\mathcal{H}({M}_{2}-{M}_{\rm gap})\mathcal{H}({M}_{1}-{M}_{2}).
\end{align}

For the sky locations, we assume sources are uniformly distributed on the celestial sphere.
In principle one should generate a 6-dimensional sample in the mass--sky-location parameters space,
but this is quite computationally intensive.
In practice, we average over a reasonable amount of sources distributed across the sky
and compress the calculation of $\langle VT\rangle$ to two (mass) dimensions.

The next term we need is $f(z,\vec{\theta},\rho^{\rm th})$, which is related to the horizon redshift of the source.
The LISA signal-to-noise of a source with frequency-domain waveform $\tilde{h}(f)$ 
at some luminosity distance
is given by~\cite{Flanagan:1997sx}
\begin{align}
{\rho}^{2} = 4\int^{{f}_{\textrm{max}}}_{{f}_{\textrm{min}}}\frac{{\tilde{{h}}^{*}(f)\tilde{h}(f)}}{{S}_{n}(f)}df,
\end{align}
where ${f}_{\textrm{min}}$ and ${f}_{\textrm{max}}$ are the initial and final frequencies.
We get the horizon redshift, and hence $f(z,\vec{\theta},{\rho}^{\textrm{th}})$, by setting $\rho={\rho}^{\textrm{th}}$.

When calculating the uncertainty and signal-to-noise for a given source, we need to integrate the waveform over a certain frequency range.
Since we are interested in sources which can in principle be detected in both LISA and the Ground, we set $f_{\rm max}\!=\!1\,{\rm Hz}$ (the conventional upper cutoff on the LISA noise curve). To determine $f_{\rm min}$, we require that a source drifts from the LISA band to the Ground band in less than a total time $T$.
The chirp time of a source with chirp mass $\mathcal{M}$ (in the observer frame) is given by~\cite{Cutler:1994ys}
\begin{align}
t = \int_{{f}_{\rm{min}}}^{f_{\rm max}}df \frac{5{c}^{5}}{96{\pi}^{8/3}} {\left(G{\mathcal{M}} \right)}^{-5/3}{{f}}^{-11/3}.
\label{chirpTime}
\end{align}
To determine $f_{\rm min}(\vec{\theta})$ we solve Eq.~\eqref{chirpTime} setting $t\equiv T$.

\begin{figure}[t]
\includegraphics[width=\columnwidth]{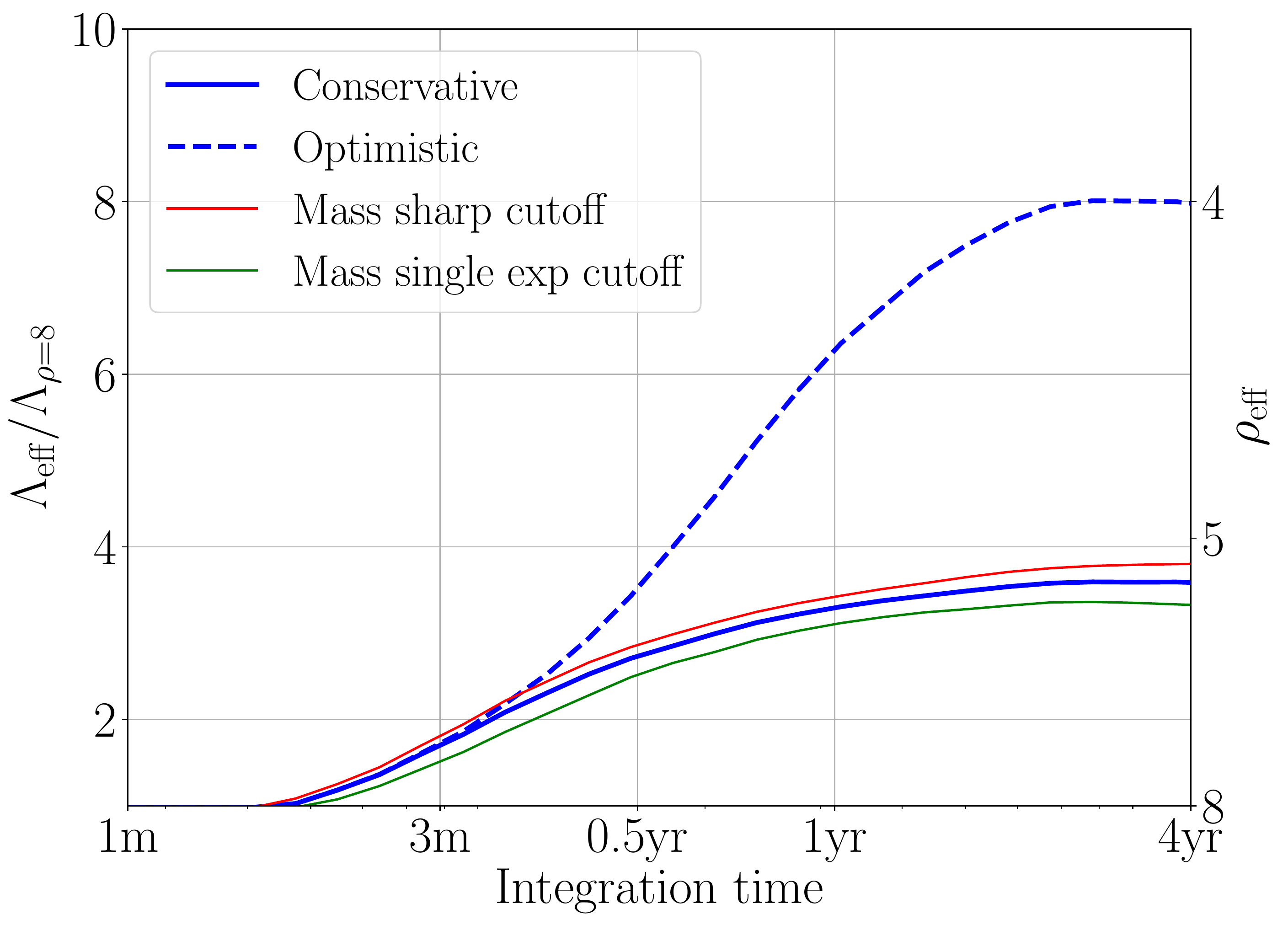}
\caption{The boost in the LISA detection rate enabled by our method, compared to setting the standard signal-to-noise threshold of $\rho=8$, and assuming that all sources are observed for the integration time $T$ given in Eq.~(\ref{chirpTime}).
The blue solid line shows the rate increase using a FAR function with a cutoff at $\rho=5$ and a mass function with a Gaussian cutoff. The dashed-blue line corresponds to a more optimistic FAR function, where the cutoff is at $\rho =4$.
For comparison, we show in red and green the result when using a mass function with a sharp cutoff  at $50{M}_{\odot}$ and a single-exponential cutoff at $40{M}_{\odot}$, respectively.}
\label{fig:FOM}
\end{figure}

In Figure~\ref{fig:FOM} we plot our main result: $\Lambda_{\rho_{\rm eff}}/\Lambda_{\rho=8}$, the increase in detection rate compared to using the standard $\rho=8$ threshold, under different assumptions.
We see that using the Ground information can boost the number of detections in LISA by a factor $\sim\!4$, under the conservative choice for the FAR. 

Since our figure-of-merit compares total rates, and we assume a constant merger rate density per comoving volume,
the uncertainties in the merger rate cancel out.
The dominant uncertainty in our result stems from the FAR. With a more optimistic choice of FAR the boost factor can increase up to $\sim8$: the LISA sampling rate~\cite{Audley:2017drz} sets a lower limit on the threshold.

The next source of uncertainty is due to the choice of mass function. The increase in detection rate is biased toward the lower end of the mass function, and so it is more significant for mass functions that favor lower mass events.
This uncertainty amounts to $\sim\!5\%$. A uniform-in-log mass function should yield similar results~\cite{Lynch:2018yom}. 

Various assumptions we have made here can be improved upon.
For example, in checking for consistency between LISA and the Ground we considered only the volume allowed by the LISA covariance matrix, instead of solving the exact two-point problem.
This is a reasonable assumption, based on the expected sensitivity of Ground observatories by the time LISA flies.

We also took the distribution of noise triggers to be uniform in the parameters of interest. 
This assumption is valid for time of coalescence and sky location,
but it may not be accurate for the two component masses.
Search template banks for ground-based detectors typically have more templates at the low-mass end~\cite{Canton:2014ena,Usman:2015kfa,Nitz:2017svb}. More realistic template banks for LISA, when available, can be used to replace the uniform distribution employed here. If the LISA and Ground templates are qualitatively similar, this replacement should increase the discarding power at the higher-mass end compared to the uniform case, and therefore improve the boost in rate.

Another approximation we made was to extrapolate our Fisher matrix calculation into the low signal-to-noise regime, where it
generally serves only as a lower bound of the uncertainties~\cite{Vallisneri:2007ev}. A more realistic estimate of the uncertainties can be achieved with other parameter estimation approaches, such as the Markov Chain Monte Carlo method~\cite{vanderSluys:2008qx}. We hope that our work will motivate participants in the ongoing LISA Data Challenges~\cite{LISADataChallenge} to verify and improve our FAR estimates.

To conclude, while the idea to use LISA detections to alert ground-based experiments about pending mergers has been explored before~\cite{Sesana:2016ljz}, we have investigated for the first time the potential of exploiting the opposite route.  

We have introduced in this {\it Letter} a method to recover sub-threshold stellar-mass BBH merger events from the LISA data stream using information from the subsequent ground-based measurements of these events. Our analysis forecasts a remarkable increase -- by a factor of $4$ to $8$, depending on the assumptions -- in  the number of LISA detections.
While our estimate was restricted to multi-band sources whose merger is detected from the Ground during the LISA lifetime,
 the same algorithm can be continuously applied for events that merge after LISA has finished its mission, yielding more detections.

The increase in number of multi-band GW detections can bring forth a plethora of rewards.
For example, improvements in parameter estimation and modeling constraints will enable novel tests of extreme gravity theories~\cite{Agathos:2013upa,TheLIGOScientific:2016src,Vitale:2016rfr,Barausse:2016eii,Yunes:2016jcc}. Most notably, discrimination between different BBH-formation channels using eccentricity~\cite{Cholis:2016kqi,Nishizawa:2016jji,Breivik:2016ddj,Nishizawa:2016eza,Gondan:2017hbp,DOrazio:2018jnv}, spins~\cite{Vitale:2015tea,Rodriguez:2016vmx,Gerosa:2017kvu,Stevenson:2017dlk,Gerosa:2018wbw}, and other waveform features~\cite{Inayoshi:2017hgw,Kremer:2018tzm,Samsing:2018ykz} will greatly benefit from the larger lever arm in frequency garnered from these measurements. 

\acknowledgments

It is our pleasure to thank Ilias Cholis, Marc Kamionkowski, Johan Samsing and Fabian Schmidt for useful discussions.
K.W.K.W. and E.B. are supported by NSF Grants No.~PHY-1841464 and AST-1841358.  
E.D.K. was supported by NASA grant NNX17AK38G.
C.C.’s work was carried out at the Jet Propulsion Laboratory, California Institute of Technology, under contract to the National Aeronautics and Space Administration. C.C. also gratefully acknowledges support from NSF Grant No.~PHY-1708212.

\bibliography{LISAHorizon}

\end{document}